\documentclass[prd,amsmath,amssymb,reprint,preprintnumbers,nofootinbib,superscriptaddress]{revtex4-1}
              
\pdfoutput=1

\usepackage{amsmath}
\usepackage{amsfonts}
\usepackage{amssymb}
\usepackage{mathrsfs}
\usepackage{graphicx}
\usepackage{color}
\usepackage[dvipsnames]{xcolor}
\usepackage{longtable}
\usepackage{bm}
\usepackage{blindtext}
\usepackage{wasysym}
\usepackage{hyperref}
\hypersetup{colorlinks=true,allcolors=blue}
\usepackage[normalem]{ulem}
\usepackage{lineno}
\usepackage{multirow,bigdelim}
\usepackage{adjustbox}
\usepackage{placeins}

\newcommand\Fontx{\fontsize{10}{12}\selectfont}

\begin{document}
 
\title{A plethora of long-range neutrino interactions probed by DUNE and T2HK\\\vskip0.2cm
\Fontx{``Contribution to the 25th International Workshop on Neutrinos from Accelerators''}}
\author{Sanjib Kumar Agarwalla}
\affiliation{Institute of Physics, Sachivalaya Marg, Sainik School Post, Bhubaneswar 751005, India}
\affiliation{Homi Bhabha National Institute, Training School Complex, Anushakti Nagar, Mumbai 400094, India}
\affiliation{Department of Physics \& Wisconsin IceCube Particle Astrophysics Center, University of Wisconsin, Madison, WI 53706, U.S.A.}
\author{Mauricio Bustamante}
\affiliation{Niels Bohr International Academy, Niels Bohr Institute, University of Copenhagen, DK-2100 Copenhagen, Denmark}
\author{Masoom Singh}
\affiliation{Institute of Physics, Sachivalaya Marg, Sainik School Post, Bhubaneswar 751005, India}
\affiliation{Department of Physics, Utkal University, Vani Vihar, Bhubaneswar 751004, India}
\author{Pragyanprasu Swain} 
\email{pragyanprasu.s@iopb.res.in (ORCID: 0000-0003-3008-480X)}
\affiliation{Institute of Physics, Sachivalaya Marg, Sainik School Post, Bhubaneswar 751005, India}
\affiliation{Homi Bhabha National Institute, Training School Complex, Anushakti Nagar, Mumbai 400094, India}

\preprint{NuFact 2024-30}

\date{\today}

\begin{abstract}
	
\begin{center} (presented by Pragyanprasu Swain)\end{center}

The next-generation neutrino oscillation experiments would be sensitive to the new neutrino interactions that would strengthen the search for physics beyond the Standard Model. In this context, we explore the capabilities of the two leading future long-baseline neutrino oscillation experiments, DUNE and T2HK, to search for new flavor-dependent neutrino interactions with electrons, protons, and neutrons that could potentially modify neutrino flavor transitions. We forecast their sensitivities in the context of long-range neutrino interactions mediated by a neutral vector boson lighter than $10^{-10}$ eV and sourced by the vast amount of
nearby and distant matter in the Earth, Moon, Sun, Milky Way, and local Universe. For the first time, we explore a plethora of  $U(1)^\prime$ symmetries inducing the new interactions built from the combination of lepton and baryon numbers. We find that in all cases, DUNE and T2HK may constrain or discover the existence of new long-range neutrino interaction, and in some favorable cases, may identify the new $U(1)^\prime$ symmetry responsible for it. In this short proceeding, we only summarize the prospects of constraining the new interaction in case of all our candidate $U(1)^\prime$ symmetries, which have been discussed in Ref.~\cite{Agarwalla:2024ylc}.       
  
\end{abstract}

\maketitle


\section{Introduction} 
\vspace{-0.25cm}
Flavor-dependent neutrino-matter interactions are promising probes of physics beyond the Standard Model. These new interactions affect $\nu_e$, $\nu_\mu$, and $\nu_\tau$ differently and thus could modify the effective Hamiltonian governing neutrino propagation, leading to observable deviations in the oscillation probabilities compared to standard oscillations.  

We explore the possibility of generating these new flavor-dependent neutrino interactions by gauging the accidental global $U(1)^\prime$ symmetries of the Standard Model (SM) involving the combination of lepton numbers ($L_e$, $L_\mu$, and $L_\tau$) and baryon number ($B$). Gauging of such a symmetry introduces a new neutral vector gauge boson, $Z^\prime$. When the mass of $Z^\prime$ is tiny ($10^{-35}-10^{-10}$ eV), the range of the new interaction becomes large, resulting in the long-range interactions (LRI) of neutrinos. These new interactions are likely to be feeble, so the modifications that they cause are difficult to spot. However, the long-range nature due to the small mediator mass helps to get the contributions to the neutrino-matter potential from local and far away celestial objects like the Earth, moon, Sun, Milky Way, and the cosmological matter distribution, resulting in a potentially appreciable modifications in the flavor transition probabilities. The highly precise detectors and well-characterized neutrino beam make long-baseline (LBL) neutrino oscillation experiments a powerful avenue for probing these sub-leading effects. We explore a plethora of $U(1)^\prime$ symmetries resulting in flavor-dependent LRI in the context of next-generation LBL experiments, DUNE~\cite{DUNE:2021cuw} and T2HK~\cite{Hyper-Kamiokande:2016srs, Hyper-Kamiokande:2018ofw}. 

In this work, we derive the sensitivity of DUNE and T2HK towards constraining the long-range interactions arising from various $U(1)^\prime$ symmetries. This contribution is based on our original publication, Ref.~\cite{Agarwalla:2024ylc}.

\vspace{-0.5cm}
\section{Long-range interaction} 
\vspace{-0.25cm}
The new neutrino interactions can be generated by extending the SM gauge group by the most general Abelian gauge symmetry, $U(1)^\prime=U(1)_{B-L}\times U(1)_{L_{\mu}-L_{\tau}}\times U(1)_{L_\mu-L_e}$, which is anomaly-free with the addition of three right-handed neutrinos to the SM particle spectrum. Any subset of the symmetry combination $c_\textsc{bl} (B-L) + c_{\mu\tau} (L_\mu-L_\tau)+ c_{\mu e} (L_\mu-L_e)$ can thus be gauged in an anomaly-free~\cite{Araki:2012ip, Heeck:2018nzc, Allanach:2018vjg, Coloma:2020gfv} way, where $a_u = a_d = c_\textsc{bl} / 3$, $a_e = b_e = -(c_\textsc{bl} +
c_{\mu e})$, $b_\mu = -c_\textsc{bl} + c_{\mu e} + c_{\mu\tau}$, and
$b_\tau = -(c_\textsc{bl} + c_{\mu\tau})$ are the $U(1)^\prime$ charges of up and down quarks, electron and electron neutrino, muon neutrino, and tau neutrino, respectively. By suitably choosing the coefficients, in this work, we consider the flavor-dependent symmetries $B-3L_e$, $L-3L_e$,  $B-\frac{3}{2}(L_\mu+L_\tau)$, $L_e-\frac{1}{2}(L_\mu+L_\tau)$, $L_e+2L_\mu+2L_\tau$, $B_y+L_\mu+L_\tau$, $B-3L_\mu$,  $L-3L_\mu$, $B-3L_\tau$,  $L-3L_\tau$, $L_e-L_\mu$, $L_e-L_\tau$, $L_\mu-L_\tau$, and $B-L_e-2L_\tau$. We define $B_y$ as $ B_1-yB_2-(3-y)B_3$~\cite{Farzan:2015doa, Coloma:2020gfv}, where $B_i$ ($i=1,2,3$) is the baryon number of $i$-th generation quark, and $y$ is an arbitrary constant. In this work, we consider the neutrino interaction with only the first generation quarks ($B_1\neq0$, $B_2=B_3=0$), hence our results are independent of the value of $y$. Each of the above symmetries, when promoted to gauge symmetries, induces flavor-dependent neutrino-matter interactions which affect $\nu_e$, $\nu_\mu$, and $\nu_\tau$ differently. Hence each symmetry will have a distinct imprint in different neutrino oscillation channels.

The neutrino propagation Hamiltonian in the presence of a long-range neutrino interaction is
\begin{equation}
\label{equ:hamiltonian_tot}
\mathbf{H}
=
\mathbf{H}_{\rm vac}
+
\mathbf{V}_{\rm mat}
+
\mathbf{V}_{\rm LRI} \;.
\end{equation}
The first term in the right-hand-side of Eq.~\ref{equ:hamiltonian_tot} is the vacuum oscillation term given by
\begin{equation}
\label{equ:hamiltonian_vac}
\mathbf{H}_{\rm vac}
=
\frac{1}{2 E}
\mathbf{U}~
{\rm diag}(0, \Delta m^2_{21}, \Delta m^2_{31})
~\mathbf{U}^{\dagger} \;,
\end{equation}
where $\mathbf{U}$ is the PMNS matrix, $E$ is the neutrino energy, $\Delta m^2_{21}$ and $\Delta m^2_{31}$ are the solar and atmospheric mass-squared differences, respectively.

The second term in the right-hand-side of Eq.~\ref{equ:hamiltonian_tot} is the standard matter contribution via the weak charged-current interaction, 
\begin{equation}
\label{equ:v_mat}
\mathbf{V}_{\rm mat}
=
{\rm diag}(V_{\rm CC}, 0, 0) \;,
\end{equation}
where $V_{\rm CC} = \sqrt{2} G_F n_e$, $G_F$ is the Fermi coupling constant, and $n_e$ is the electron number density along the neutrino path. 

The last term in the right-hand-side of Eq.~\ref{equ:hamiltonian_tot} is the contribution from the new interaction,
\begin{equation}
\label{equ:lri_pot}
\mathbf{V}_{\rm LRI}
=
{\rm diag}(V_{{\rm LRI}, e}, V_{{\rm LRI},\mu}, V_{{\rm LRI},\tau}) \;.
\end{equation}
Here, $V_{\rm LRI, \alpha}~(\alpha=e,\mu,\tau)$ denotes the long-range interaction (LRI) potential experienced by the neutrino of flavor $\alpha$. Depending on the specific symmetry, the entries of the LRI potential matrix could take non-trivial values. Their values depend on the $U(1)^\prime$ charges of matter fermions and various neutrino flavors and also on the nature of matter distribution inside any particular celestial body. The textures of the LRI matrices induced by all our candidate symmetries are discussed in detail in Ref.~\cite{Agarwalla:2024ylc}; see Table 1 there. 

\vspace{-0.5cm}
\section{Results}
\vspace{-0.25cm}
Having described the detailed statistical methods in our primary publication (see Ref.~\cite{Agarwalla:2024ylc}), where we define the test statistic for sensitivity to the long-range potential as a Poissonian $\Delta \chi^2$ and minimize it over the most uncertain oscillation parameters, $\theta_{23}$ and $\delta_{\rm CP}$ and the neutrino mass ordering, we present in this section one of the main focus of our work, the prospects of constraining the LRI potential induced by each of the candidate $U(1)^\prime$ symmetries. See Table~\ref{tab:params_value1} for the oscillation parameter values and their minimization ranges considered in the fit while deriving the constraints.

\begin{table}
	\centering
	\begin{center}
		\begin{adjustbox}{width=\linewidth} 
			\renewcommand{\arraystretch}{1.6}
			\begin{tabular}{|c|c|c|c|}
				\hline
				Parameter &  Best-fit value & 3$\sigma$ range &  Statistical treatment \\
				\hline
				${\theta_{12}}~[^{\circ}]$ & 33.45 & 31.27--35.87  & Fixed to best fit \\
				\hline
				{${\theta_{13}}~[^{\circ}]$} & 8.62 & 8.25--8.98 & {Fixed to best fit} \\ 
				\hline
				{${\theta_{23}}~[^{\circ}]$} & 42.1 & 39.7--50.9 & {Minimized over $3\sigma$ range} \\
				\hline
				{$\delta_{\rm CP}~[^{\circ}]$} & 230 & 144--350 & {Minimized over $3\sigma$ range} \\
				\hline
				{$\frac{\Delta{m^2_{21}}}{10^{-5} \, \rm{eV}^2}$} & {7.42} & {6.82--8.04}  & {Fixed to best fit} \\
				\hline
				{$\frac{\Delta{m^2_{31}}}{10^{-3} \, \rm{eV}^2}$} & 2.51  & 2.430--2.593 & {Minimized over $3\sigma$ range} \\
				\hline
			\end{tabular}
		\end{adjustbox}
		\caption{\textit{Best-fit values and $3\sigma$ allowed ranges of the oscillation parameters used in our analysis.} The values are taken from the NuFIT~5.1 global fit to oscillation data~\cite{Esteban:2020cvm, NuFIT}.  The true neutrino mass ordering is assumed to be normal for all the results presented in this article.}
		\label{tab:params_value1}
	\end{center}
\end{table}
\begin{figure}
	\centering
	\includegraphics[width=1\columnwidth]{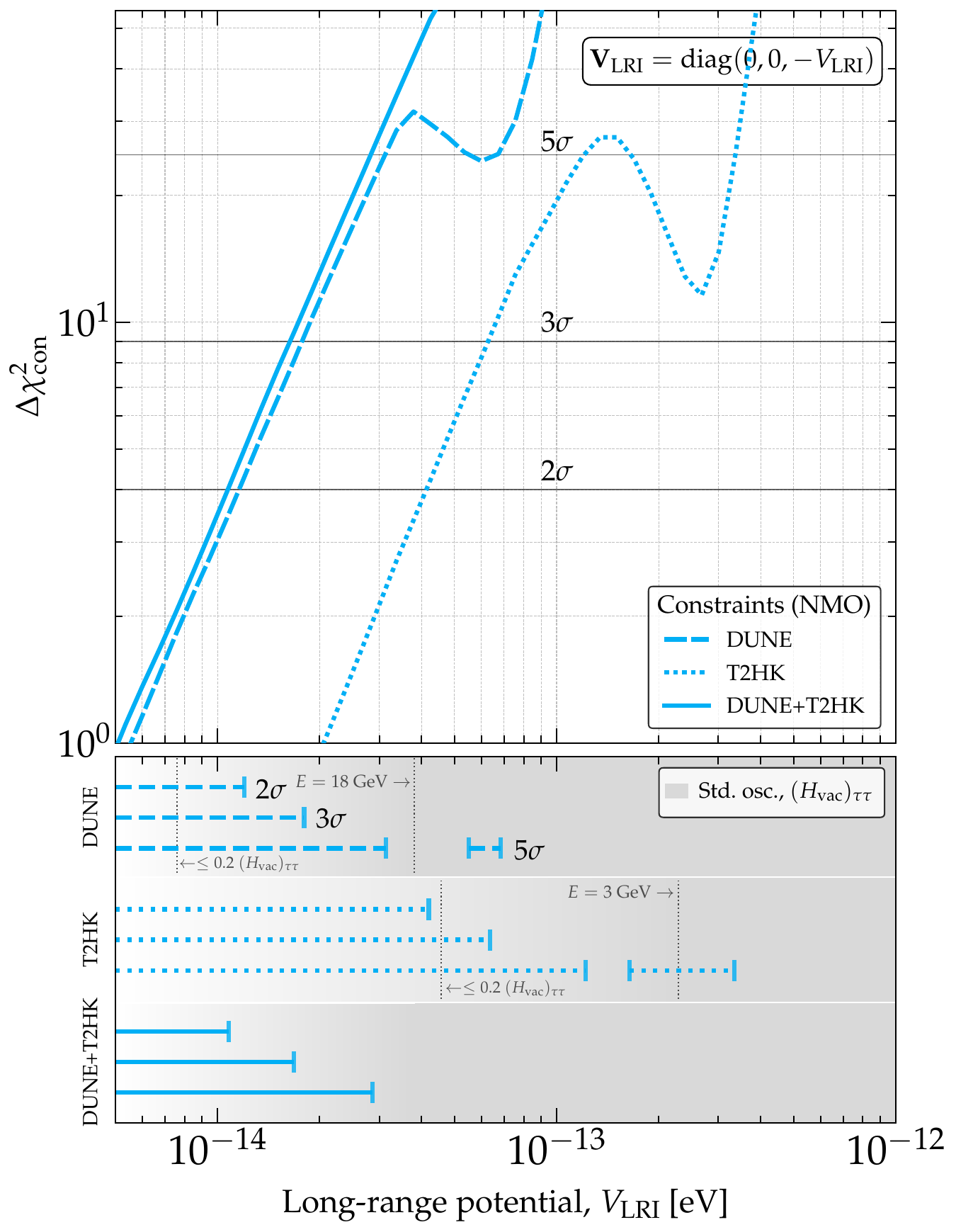}
	\caption{\textit {Projected test statistic used to constrain the new matter potential induced by a $U(1)^\prime$ symmetry.} Here, we show the limits for an illustrative texture of the LRI potential, $\mathbf{V}_{\rm LRI} = \textrm{diag}(0, 0, -V_{\rm LRI})$ induced by symmetries $L - 3L_\tau$ or $B - 3L_\tau$. The true neutrino mass ordering is assumed to be normal. The experiments are sensitive to values of $V_{\rm LRI}$ that are comparable to the standard-oscillation terms in the Hamiltonian; for the choice of $\mathbf{V}_{\rm LRI}$ texture in this figure, this is $(\mathbf{H}_{\rm vac})_{\tau\tau}$, which scales as $1/E$. Constraints on $V_{\rm LRI}$ lie around 20\% of the value of $(\mathbf{H}_{\rm vac})_{\tau\tau}$ estimated at the highest accessible energy in each experiment.  
		This figure is taken from Ref.~\cite{Agarwalla:2024ylc}.}
	\label{fig:lrf_bounds_NH_selective}
\end{figure} 

\begin{figure*}
	\centering
	\includegraphics[width=0.781\linewidth]{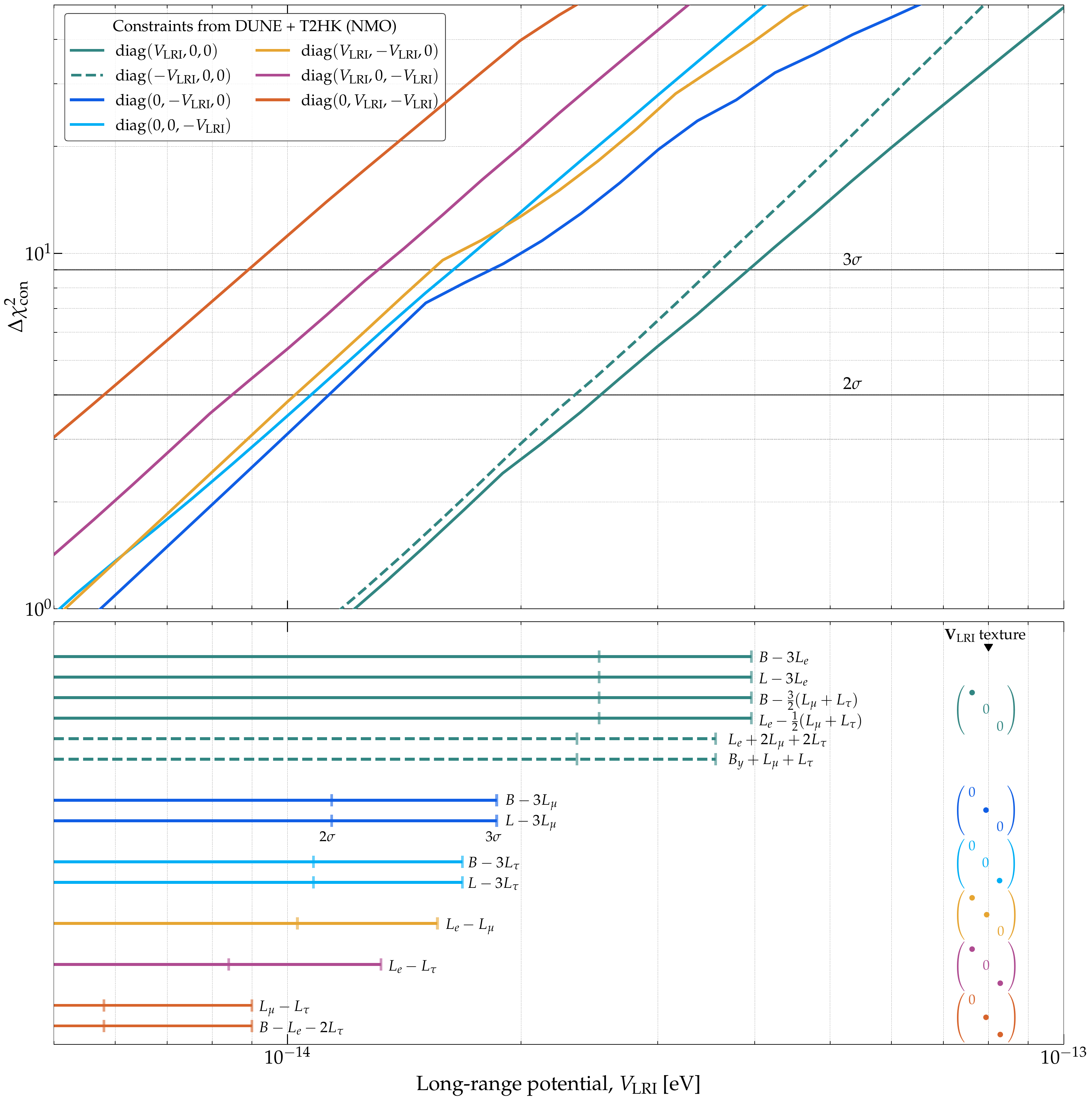}
	\caption{\textit{Projected test statistic (top) used to derive upper limits (bottom) on the new matter potential arising from our candidate $U(1)^\prime$ symmetries.}  The true neutrino mass ordering is assumed to be normal. Results are obtained from the combined analysis of prospective data from DUNE and T2HK. For details on the texture of the new neutrino-matter interaction potential, $\mathbf{V}_{\rm LRI}$, refer to Table 1 of Ref.~\cite{Agarwalla:2024ylc}.
		\label{fig:constraints_on_pot_dune_t2hk-NMO}}
\end{figure*}
Figure~\ref{fig:lrf_bounds_NH_selective} shows the test statistics used to place constraints on the LRI potential for standalone DUNE, T2HK, and their combination DUNE+T2HK, for a particular texture of LRI potential, $\mathbf{V}_{\rm LRI}={\rm diag}(0,0, -V_{\rm LRI})$. We can see from the upper panel that there exist dips in the test statistics in case of the individual experimental setups. This is attributed to the degeneracies between the LRI potential, $V_{\rm LRI}$, and the most uncertain standard oscillation parameters, $\theta_{23}$ and $\delta_{\rm CP}$, and the choice of neutrino mass ordering. The dip in the test statistic in the case of DUNE by itself, around $V_{\rm LRI} \sim 5\times10^{-14}$ eV, represents a weakening of the constraining power. It is mainly due to the degeneracy between $V_{\rm LRI}$, $\theta_{23}$, and $\delta_{\rm CP}$ and in case of T2HK it is the degeneracy between $V_{\rm LRI}$ and the mass ordering. Combining both the experiments removes the intrinsic degenerate solutions enhancing the sensitivity to the LRI potential.  

Figure~\ref{fig:constraints_on_pot_dune_t2hk-NMO} shows the projected constraints for all the candidate symmetries considered in this work. We find that {\it regardless of which symmetry is responsible for inducing the new interaction, DUNE and T2HK are able to constrain the new interaction to a level comparable to standard-oscillation terms in the Hamiltonian ($\sim 10^{-14}-10^{-13} ~{\rm eV}$)}. The symmetries with new matter potential containing non-zero entries in the $\mu-\tau$ sector affect mostly the $\nu_\mu\to\nu_\mu$ disappearance probability, and those containing non-zero $ee$-entry affect primarily the $\nu_\mu\to\nu_e$ appearance probability. We observe that the symmetries that primarily affect the $\nu_\mu\to\nu_\mu$ disappearance channel get the tightest limits due to high disappearance event rates in DUNE and T2HK and the ones that predominantly influence the $\nu_\mu\to\nu_e$ appearance channel get the weakest constraints due to less appearance event counts in the considered experiments.    

\begin{figure}
	\centering
	\includegraphics[width=1.0\linewidth]{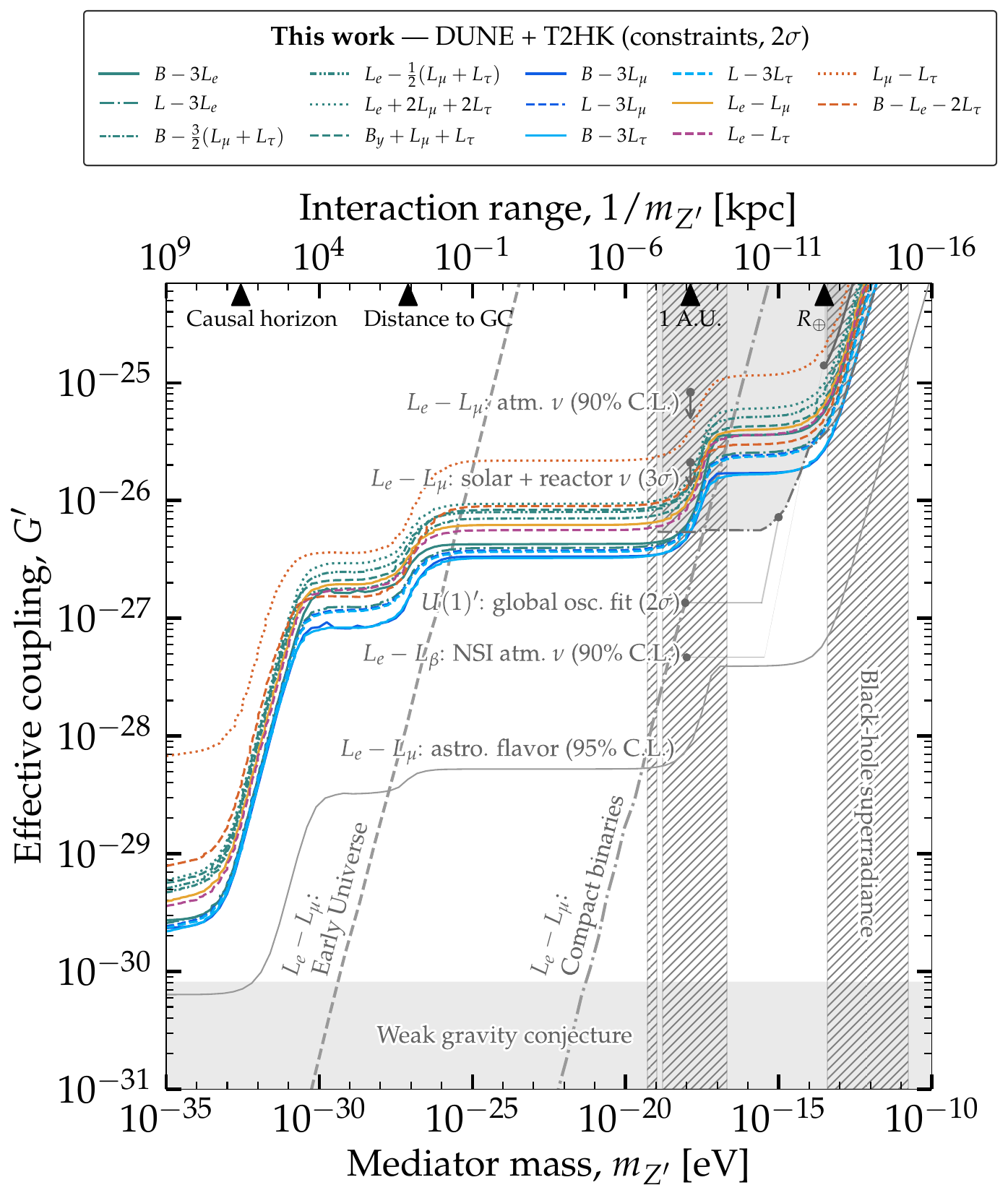}
	\caption{\textit{Projected upper limits on the effective coupling of the new gauge boson, $Z^\prime$, that mediates flavor-dependent long-range neutrino interactions.}  Results are for combined setup DUNE+T2HK , after 10 years of operation, and for each of our candidate $U(1)^\prime$ symmetries.  For this figure, we assume that the true neutrino mass ordering is normal. For each symmetry, the limits on the coupling, $G^\prime$, as a function of the mediator mass, $m_{Z^\prime}$, are converted from the limits on $V_{\rm LRI}$ in fig.~\ref{fig:constraints_on_pot_dune_t2hk-NMO}. Existing limits are shown in gray curves and gray-shaded regions. For details on existing limits, see Ref.~\cite{Agarwalla:2024ylc}.}
	\label{fig:all_symmetry}
\end{figure}

Figure~\ref{fig:all_symmetry} shows the constraints on the coupling vs. mass plane of the new neutral gauge boson, $Z^\prime$. These constraints are derived from the upper limits on $V_{\rm LRI}$ (for the detailed procedure, see Ref.~\cite{Agarwalla:2024ylc}). Because different celestial objects have a different abundance of electrons, protons, and neutrons, the tightest constraints on the LRI potential in Fig.~\ref{fig:constraints_on_pot_dune_t2hk-NMO} do not necessarily translate into the tightest constraints on the coupling strength in Fig.~\ref{fig:all_symmetry}. Moreover, we find that {\it regardless of which symmetry is responsible for inducing the long-range interaction, DUNE and T2HK may outperform the existing limits on the coupling strength, especially for the mediators lighter than $10^{-18}$ eV.} Though it seems that the limits from the astrophysical flavor measurements~\cite{Agarwalla:2023sng} are stronger than ours, but they are limited by the large astrophysical uncertainties that are not taken into account. 
\vspace{0.15cm}
\section{Conclusions}
\vspace{-0.25cm}
The high precision measurements of the neutrino oscillation parameters make the oscillation experiments sensitive to tiny beyond the Standard Model (BSM) effects. In this study, we focus on one such BSM scenario: the long-range interactions (LRI) of neutrinos. We explore a plethora of $U(1)^\prime$ symmetries that result in new flavor-dependent neutrino-matter interactions. With ten years of exposure from each of the next-generation long-baseline experiments, DUNE and T2HK, we derive the projected constraints on the LRI potential in case of each of the candidate symmetry and also extend these constraints to the mass vs. coupling plane of the new neutral mediator, $Z^\prime$, responsible for inducing LRI. {\it We find that regardless of which symmetry is responsible for inducing the new long-range neutrino-matter interactions, DUNE and T2HK may constrain them more strongly than ever before, provided the new matter potential is comparable to that of the standard oscillation scale.}

\vspace{-0.35cm}
\begin{acknowledgments}
\vspace{-0.2cm}
We acknowledge financial support from DAE, DST, DST-SERB, Govt. of India, INSA, and USIEF. M.B.~is supported by the {\sc Villum Fonden} under the project no. 29388. The numerical simulations 
are performed using “SAMKHYA: High-Performance Computing Facility” 
provided by the Institute of Physics, Bhubaneswar, India.
\end{acknowledgments}

\bibliographystyle{apsrev4-1}
\bibliography{refer-lri.bib}

\end{document}